\documentstyle[preprint,aps,epsfig]{revtex}
\tightenlines
\begin{document}

\title{Is Cosmic Acceleration Really Recent?\footnote{To 
appear in proceedings of "Cosmology and Elementary Particle Physics", Coral
Gables Conference, December 2001, B. N.  Kursunoglu (Ed.), American Institute of
Physics, NY (2002).}}

\author{\normalsize{Philip D. Mannheim\footnote{Email address: 
mannheim@uconnvm.uconn.edu}} \\
\normalsize{Department of Physics,
University of Connecticut, Storrs, CT 06269} \\}

\maketitle

\begin{abstract}
In the standard cosmological paradigm cosmic acceleration is to only be a 
very recent (viz. $z \leq 1$) phenomenon, with the universe being required
to be decelerating at all higher redshifts. We suggest that this particular
expectation of the standard model is to be viewed as a quite definitive
test not only of the model itself but also of the fine-tuning assumption on
which the expectation is based, with the expectation itself actually being
readily amenable to testing once the Hubble plot can be extended out to
only $z=2$ or so. Moreover, such a modest extension of the Hubble plot will
also provide for definitive testing of the non fine-tuned alternate
conformal gravity theory, a theory in which the universe is to accelerate
both above and below $z=1$.  
\end{abstract}

\section{The Hubble plot of standard cosmology}
\medskip

In a standard pure matter or pure radiation Friedmann cosmology the 
attractive nature of gravity entails the existence of an initial big bang
singularity followed by a subsequent decelerating expansion. Primary
evidence in general favor of such a picture is obtained from observational
study of three widely separated epochs, viz. early universe
nucleosynthesis, the recombination era cosmic microwave background, and the
current $z\leq 1$ era $(d_L,z)$ Hubble plot. While it had long been thought
that a decelerating expansion was to occur in every epoch, data accumulated
only recently
\cite{Riess1998,Perlmutter1999,Bahcall2000,deBernardis2000} now reveal the
presence of an unanticipated additional repulsive component to cosmological
gravity, a component most commonly attributed to the presence of a 
non-vanishing cosmological constant $\Lambda$, a component whose
contribution to cosmic evolution is only found to start to become of
consequence at around $z=1$ or so where its presence is then central to the
elucidation of the $z\leq 1$ Hubble plot data.\footnote{While the cosmic
microwave background data \cite{deBernardis2000} are certainly compatible
with the presence of a non-vanishing $\Lambda$, in and of themselves alone
they can just as readily support a universe in which $\Lambda$ is absent.}
Specifically, through use of the standard Einstein-Friedmann cosmological
evolution equation	
\begin{equation}
\dot{R}^2(t) +kc^2=\dot{R}^2(t)(\Omega_{M}(t)+\Omega_{\Lambda}(t))
\label{1}
\end{equation}
[where $\Omega_{M}(t)=8\pi G\rho_{M}(t)/3c^2H^2(t)$ is due to ordinary 
$\rho_M(t)\sim 1/R^n(t)$ matter and where $\Omega_{\Lambda}(t)=8\pi
G\Lambda/3cH^2(t)$ is due to a cosmological constant $c\Lambda$]
phenomenological data fitting has been found to yield current era values of
$0.3$ for $\Omega_{M}(t_0)$ and $0.7$ for $\Omega_{\Lambda}(t_0)$, to thus
entail that in the current era the deceleration parameter
$q_0=q(t_0)=(n/2-1)\Omega_{M}(t_0) -\Omega_{\Lambda}(t_0)$ has to take the
negative value of $-1/2$, with the current era universe thus not being a
decelerating one after all. 

While the identifying of such specific values for $\Omega_{M}(t_0)$, 
$\Omega_{\Lambda}(t_0)$ and $\Omega_{k}(t_0)=-kc^2/\dot{R}^2(t_0)
=1-\Omega_{M}(t_0) -\Omega_{\Lambda}(t_0)$ has enabled standard cosmology to
ostensibly achieve its primary purpose, namely that of determining the 
matter, vacuum and curvature content of the universe, and while the obtained
values even provide support for the flat $\Omega_{k}(t)=0$ inflationary
universe paradigm \cite{Guth1981}, the particular values obtained for these
parameters are nonetheless extremely perplexing. Specifically, a priori
estimates for $c\Lambda\equiv \sigma T_V^4$ would suggest for $T_V$ a value
of either a typical particle physics temperature scale of order $10^{16}$
degrees or so or a quantum-gravitational Planck temperature scale of order
$10^{33}$ degrees, to thereby yield an $\Omega_{\Lambda}(t_0)
/\Omega_{M}(t_0)$ ratio of order $10^{60}$ to $10^{120}$, a ratio not only
overwhelmingly larger than the requisite measured value of order one, but
one which would (for an $\Omega_{M}(t_0)$ of order one) entail that
$\Omega_k(t_0)$ would have to be of order $-10^{60}$ and thus be nowhere
near flat at all.

To get round this problem the standard paradigm thus proposes that rather 
than use such a fundamental physics based $T_V$ one should instead, and
despite the absence of any currently known justification, fine-tune $T_V$
down by orders and orders of magnitude so that the value
$\Omega_{\Lambda}(t_0)=0.7$ would then ensue. Beyond the difficulty
inherent in trying to understand how this might actually be dynamically
achieved, even a successful resolution of this issue would still not
actually leave cosmology totally free of fine-tuning problems, since in its
turn, having a current era $\Omega_{\Lambda}(t_0)$ of order one creates a
yet further problem for the standard model, one of then having to have an
expressly fine-tuned early universe. Specifically, for an
$\Omega_{\Lambda}(t_0)$ of order one the early universe associated with Eq.
(\ref{1}) would need to be one in which $\Omega_{M}(t=t_{PL})$ would have 
had to have been incredibly close to one at the Planck time $t=t_{PL}$,
while $\Omega_{\Lambda}(t=t_{PL})$ itself would have had to have been as
small as $O(10^{-120})$. The early universe thus has only a one in
$10^{120}$ chance of ever evolving into our current universe unless some
explicit dynamical mechanism could be found which would naturally fix these
needed initial conditions with incredible precision. In a sense this
fine-tuning problem is a new variant of the venerable flatness problem. As
we recall, in the pre $\Omega_{\Lambda}(t_0) \neq 0$ days it was very
difficult to understand why an $\Omega_{M}(t)$ which had been redshifting
for more than 10 billion years would be anywhere near one today rather than
being orders and orders of magnitude smaller, with it being inflation which
then provided a natural answer to this problem. Specifically, it was shown
by Guth \cite{Guth1981} that if there were to be a period of rapid de
Sitter inflation (viz. rapid acceleration) prior to the onset of the
current Robertson-Walker (RW) era, then such an inflationary era would
precisely lead at its end to a set of initial conditions for an ensuing RW
phase in which $\Omega_{M}(t)$ would not merely be close to one but would in
fact be identically equal to one in each and every epoch and thus not
susceptible to redshifting at all. However, with the advent of a non-zero
cosmological constant inflation now only fixes the sum of $\Omega_{M}(t)$
and $\Omega_{\Lambda}(t)$ to be equal to one in all epochs but does not
constrain their ratio. Currently then, standard cosmology stands waiting
for the development of some sort of generalized version of early universe
inflation which would naturally lead to initial RW era conditions which
then would naturally fix the initial values of both $\Omega_{M}(t)$ and
$\Omega_{\Lambda}(t)$ to the requisite precision. This then is the
challenge to the standard cosmology posed by the new $z\leq 1$ Hubble plot
data.

As regards actually fitting these Hubble plot data, we note that when viewed
purely as a phenomenology (i.e. without regard to any of the above 
fine-tuning concerns) an $\Omega_{M}(t_0)=0.3$, $\Omega_{\Lambda}(t_0)=0.7$
standard model then performs extraordinarily well. Through use of type Ia
supernovae as standard candles the authors of \cite{Riess1998} and
\cite{Perlmutter1999} were able to extend the Hubble plot of luminosity
versus redshift out to redshifts close to one. To illustrate the quality of
the fits which then ensue we follow \cite{Perlmutter1999} and fit 38 of
their reported 42 data points together with 16 of the 18 earlier lower $z$
points of \cite{Hamuy1996}, for a total of 54 data points with reported
effective blue apparent magnitude $m_i$ and uncertainty $\sigma_i$. (While
we thus, following \cite{Perlmutter1999}, leave out 6 questionable data
points for the fitting, nonetheless, for completeness we still include them
in the displayed Fig. (1).) For the fitting we calculate the apparent
magnitude $m$ of each supernova at redshift $z$ via $m=25+M+5log_{10} d_L$
(the luminosity distance $d_L$ being in Mpc) where $M$ is their assumed
common absolute magnitude, and find for $\Omega_{M}(t_0)=0.3$,
$\Omega_{\Lambda}(t_0)=0.7$ and $M=-19.37$ that $\chi^2=\sum
(m-m_i)^2/\sigma_i^2$ takes the value 57.74, with the fit itself being
displayed as the lower curve in Fig. (1). \footnote{The very fact that
the data can be fitted so well with a common $M$ at all (and even with one
of value typical of nearby supernovae) strongly suggests that type Ia
supernovae are indeed good standard candles.} As the fitting shows, once
one allows for the gravitational repulsion associated with a non-vanishing
$\Omega_{\Lambda}(t_0)$ the standard model can nicely account for the 
supernovae data.

With the values of $\Omega_{M}(t_0)=0.3$, $\Omega_{\Lambda}(t_0)=0.7$ thus 
being established by the $z\leq 1$ Hubble plot data, we now note that since
the matter density $\rho_{M}(t)$ redshifts while $\Lambda$ of course does
not, as we go to higher and higher redshift the
$\Omega_{M}(t)/\Omega_{\Lambda}(t)$ ratio will get bigger and bigger, with
the attractive matter density numerically being found to overcome the
repulsive $\Lambda$ contribution to the deceleration parameter at a
redshift of only $z=0.67$. In the standard model then the universe would be
such that it would decelerate ($q(t)>0$) continually in all epochs until
the matter density contribution finally manages to redshift itself down to
the cosmological constant contribution, something which is to occur at the
incredibly late $z=0.67$ when $q(t)$ would at long last finally change sign.
Indeed, the particular timing of this change in sign is itself a reflection
of the standard model early universe fine-tuning problem we discussed
earlier, with initial conditions having to be such that this change over
would occur precisely in our own epoch, neither earlier than it nor later.
Now while it is very peculiar that such a turn around is to occur just in
our own particular epoch, nonetheless, independent of one's views regarding
the merits or otherwise of such an expectation, the prediction itself is
actually readily amenable to testing, with just a modest increase in the
range of $z$ (say to $z=2$ or so) in the Hubble plot being able to reveal
its possible presence. Moreover, such a study would be a completely
kinematic one, one totally independent of dynamical assumptions (such as
those for instance required for the extraction of cosmological parameters
from the cosmic microwave background) and would thus be completely clear
cut. In this sense then study of the $z>1$ Hubble plot can provide a
completely dynamics independent test of whether or not $\Lambda$ really is
as small as the standard model's assumed fine-tuning would require. With
the $z>1$ Hubble plot thus being the "smoking gun" for a fine-tuned
$\Lambda$, we thus exhibit in Fig. (2) the standard model expectation (the
lowest curve) out to $z=5$.  In and of itself then it would be extremely
informative to extend the range of the Hubble plot. However, as we now show, 
it would be of additional interest since it would allow for a rather 
unequivocal comparison between standard cosmology and the recently proposed
alternate conformal cosmology, a theory which is capable of fitting the
very same supernovae data without any fine-tuning at all.

\section{The Hubble plot of conformal cosmology}

Given both the fine-tuning problems of the standard cosmology and the 
absence to date of any solution to them, it is thus of value to entertain
and explore possible candidate alternate cosmologies to see if any one of
them might shed some light on the issue. Now while the choice of possible
alternate theories is quite vast (pure metric based theories of gravity
require only a general coordinate scalar action, of which there is an
infinite number containing derivatives of the Riemann tensor out to
arbitrarily high order), one particular such alternative is explicitly
singled out. Specifically, since it possesses a symmetry which when
unbroken obliges the cosmological constant to vanish identically
\cite{Mannheim1990}, conformal gravity (viz. gravity based on the fully
covariant, locally conformal invariant Weyl action 
\begin{equation}
I_W=-\alpha_g \int d^4x
(-g)^{1/2} C_{\lambda\mu\nu\kappa}  C^{\lambda\mu\nu\kappa}
\label{2}
\end{equation}
where $C^{\lambda\mu\nu\kappa}$ is the conformal Weyl tensor and where
$\alpha_g$ is a purely dimensionless gravitational coupling constant) is
immediately suggested and motivated. The cosmology associated with the 
conformal gravity theory was first presented in \cite{Mannheim1992} where it
was shown to both possess no flatness problem (to thereby release conformal
cosmology from the need for the copious amounts of cosmological dark matter
required of the standard model) and to have an effective cosmological
Newton constant, $G_{eff}$, which actually turned out to be negative.  Thus
long in advance of the recent supernovae data it had been noted that
conformal cosmology possessed a repulsive gravitational
component.\footnote{In fact, equally in advance of the supernovae data, it
had also been noted \cite{Mannheim1996} that in a $\Lambda=0$ conformal
cosmology the current era $q_0$ would then be identically equal to zero,
with a $\Lambda=0$ conformal cosmology thus possessing a repulsion not
present in a $\Lambda=0$ standard cosmology where $q_0=1/2$.} Subsequently
\cite{Mannheim1996,Mannheim1998}, the cosmology was shown to also possess no
horizon problem or universe age problem. And finally, it was shown
\cite{Mannheim2000,Mannheim2001a} that even after the conformal symmetry is
spontaneously broken by a $\Lambda$ inducing scale breaking cosmological 
phase transition, the theory continues to be able to keep the contribution
of the induced cosmological constant to cosmic evolution under control even 
in the event that $\Lambda$ is in fact as big as particle physics suggests,
to thereby provide a completely natural solution to the cosmological
constant problem without the need for any fine tuning at all. In the
present paper we use the results of \cite{Mannheim2000,Mannheim2001a} to
show that conformal gravity not only controls the cosmological constant in
principle, in practice it even provides for a completely acceptable
accounting of the recent supernovae Hubble plot data as well.   

Analysis of the implications of conformal cosmology is greatly facilitated 
by considering the generic conformal matter action
\begin{equation}
I_M=-\hbar\int d^4x(-g)^{1/2}[S^\mu S_\mu/2 -
S^2R^\mu_{\phantom{\mu}\mu}/12 + \lambda S^4 +
i\bar{\psi}\gamma^{\mu}(x)(\partial_\mu + \Gamma_\mu(x))\psi -
gS\bar{\psi}\psi] 
\label{3}
\end{equation}
for massless fermions and a conformally coupled order parameter scalar 
field. When the scalar field breaks the conformal symmetry by acquiring a
non-zero expectation value $S_0$, the energy-momentum tensor associated
with the matter action of Eq. (\ref{3}) is found (for a perfect matter fluid
$T^{\mu\nu}_{kin}$ of the fermions) to take the form \cite{Mannheim2001a}        
\begin{equation}
T^{\mu\nu}=T^{\mu\nu}_{kin}-\hbar S_0^2(R^{\mu\nu}-
g^{\mu\nu}R^\alpha_{\phantom{\alpha}\alpha}/2)/6            
-g^{\mu\nu}\hbar\lambda S_0^4~~,
\label{4}
\end{equation}
with the complete solution to the scalar, fermionic and gravitational field
equations of motion in a background RW geometry (viz. a geometry in which
$C^{\lambda\mu\nu\kappa}=0$) then reducing
\cite{Mannheim2001a} to the remarkably simple equation $T^{\mu\nu}=0$, i.e.
reducing to
\begin{equation}
\hbar S_0^2(R^{\mu\nu}-
g^{\mu\nu}R^\alpha_{\phantom{\alpha}\alpha}/2)/6=T^{\mu\nu}_{kin}           
-g^{\mu\nu}\hbar\lambda S_0^4~~,
\label{5}
\end{equation}
with the vanishing of $T^{\mu\nu}$ immediately fixing the zero of energy. 
As we thus see, the evolution equation of conformal cosmology looks
identical to that of standard gravity save only that the quantity $-\hbar
S_0^2 /12$ has replaced the familiar $c^3/16 \pi G$, so that instead of
being attractive the effective cosmological $G_{eff}=-3c^3/4\pi \hbar
S_0^2$ is actually negative, and instead of being fixed as the standard low
energy Newtonian $G$, the cosmological $G_{eff}$ is instead fixed by the
altogether different scale $S_0$, a scale which when large enough would
yield an effective  cosmological $G_{eff}$ which would then be altogether
smaller than the standard Cavendish $G$.\footnote{In fact, with the
non-relativistic terrestrial and solar system conformal gravity
expectations being controlled \cite{Mannheim1994} by a local $G$ whose
dynamical generation is totally decoupled
\cite{Mannheim2000,Mannheim2001a} from that of the cosmological
$G_{eff}$, in conformal gravity cosmology is thus completely freed
from the need to be controlled by the Cavendish $G$.} 

Given the equation of motion of Eq. (\ref{5}), the ensuing conformal 
cosmology evolution equation is then found (on setting $\Lambda=\hbar\lambda
S^4_0$) to take a form remarkably similar to that of Eq. (\ref{1}), viz. 
\begin{eqnarray}
\dot{R}^2(t) +kc^2
=\dot{R}^2(t)(\bar{\Omega}_{M}(t)+
\bar{\Omega}_{\Lambda}(t)) 
\label{6}
\end{eqnarray}
where $\bar{\Omega}_{M}(t)=8\pi G_{eff}\rho_{M}(t)/3c^2H^2(t)$,
$\bar{\Omega}_{\Lambda}(t)=8\pi G_{eff}\Lambda/3cH^2(t)$. Further, unlike 
the situation in the standard theory where preferred values for the relevant
evolution parameters (such as the magnitude and even the sign of $\Lambda$) 
are only determined by the data fitting itself, in conformal gravity
essentially everything is already a priori known. With conformal gravity
not needing dark matter to account for galactic rotation curve systematics
\cite{Mannheim1997}, $\rho_{M}(t_0)$ can be determined directly from
luminous matter alone, with galaxy luminosity counts giving a value for it
of  order $0.01\times 3c^2H^2_0/8\pi G$ or so. Further, with $c\Lambda$
being generated by an energy density lowering particle physics vacuum
breaking phase transition in an otherwise scaleless theory, $c\Lambda$ (and
thus the $\hbar\lambda S_0^4$ term which simulates it) must unambiguously
be negative, with it thus being typically given by $-\sigma T_V^4$ where
$T_V$ is a necessarily particle physics sized scale. Then with $G_{eff}$
also being negative, the quantity $\bar{\Omega}_{\Lambda}(t)$ itself must
thus be positive, just as needed to give cosmic acceleration
($q(t)=(n/2-1)\bar{\Omega}_{M}(t)-\bar{\Omega}_{\Lambda}(t)$). Similarly,
the sign of the spatial 3-curvature $k$ is known from theory
\cite{Mannheim2000} to be negative, something which has been independently
confirmed from a phenomenological study of galactic rotation curves
\cite{Mannheim1997}. Moreover, since $G_{eff}$ is negative, the cosmology is
singularity free and thus expands from a (negative curvature supported) 
finite maximum temperature $T_{max}$, a temperature which is necessarily
greater \cite{Mannheim1998,Mannheim2000} (and potentially even much greater
\cite{Mannheim2001a}) than $T_V$. And finally, with $G_{eff}$ being negative,
the quantity $\bar{\Omega}_M(t)$ must be negative for ordinary 
$\rho_{M}(t)>0$ matter, with $q(t)$ thus being negative in all
epochs.\footnote{Included in this  class of $q(t)<0$ universes are the
coasting ones in which $q(t)=0^{-}$.} Consequently in the conformal theory
we never need to fine tune in order to make any particular epoch such as
our own be an accelerating one, with repulsive cosmological gravity thus
being completely natural to conformal gravity in each and every epoch.    

Given only that $\Lambda$, $k$ and $G_{eff}$ are in fact all negative in the
conformal theory, the evolution of the theory is then completely determined,
with the expansion rate parameters being found
\cite{Mannheim1998,Mannheim2000,Mannheim2001a} to be given by
\begin{equation}
R^2= -k(\beta-1)/2\alpha
-k\beta \sinh^2 (\alpha^{1/2} ct)/\alpha~~,~~T_{max}^2/T^2=
1+2\beta \sinh^2 (\alpha^{1/2} ct)/(\beta-1)~~,
\label{7}
\end{equation}
where  
$\beta=(1- 16A\lambda/k^2\hbar c)^{1/2}=(1+T_V^4/T_{max}^4)/
(1-T_V^4/T_{max}^4)$, $\alpha c^2=-2\lambda S_0^2=8\pi G_{eff}\Lambda/3c$.
In terms of the parameters $T_{max}$ and $T_V$ we thus obtain
\begin{eqnarray}
\tanh^2 (\alpha^{1/2} ct)=(1-T^2/T_{max}^2)/(T_{max}^2T^2/T_V^4+1)~~,
\nonumber \\
H(t)=\alpha^{1/2}c(1-T^2/T^2_{max})/\tanh(\alpha^{1/2}ct)~~,
\nonumber \\
\bar{\Omega}_{\Lambda}(t)= 
(1-T^2/T_{max}^2)^{-1}(1+T^2T_{max}^2/T_V^4)^{-1},~~ 
\bar{\Omega}_M(t)=-(T^4/T_V^4)\bar{\Omega}_{\Lambda}(t)
\label{8}
\end{eqnarray}
at any $T(t)$ without any approximation at all. From Eq. (\ref{8}) we now
see that simply because $T_{max}$ is overwhelmingly larger than the current
temperature $T(t_0)$, i.e. simply because the universe is as old as it is,
it automatically follows, without any fine-tuning at all, that the current
era $\bar{\Omega}_{\Lambda}(t_0)$ has to lie somewhere between zero and one
today no matter how big (or small) $T_V$ might actually be, with conformal
gravity thus having total control over the contribution of the cosmological
constant to cosmic evolution. Conformal gravity thus solves the
cosmological constant problem by quenching $\bar{\Omega}_{\Lambda}(t_0)$
rather than by quenching $\Lambda$ itself (essentially by having a
$G_{eff}$ which is altogether smaller than the standard $G$), and with it
being the quantity $\bar{\Omega}_{\Lambda}(t_0)$ which is the one which is
actually measured in cosmology, it is only its quenching which is actually
needed. With conformal gravity thus being able to naturally accommodate a
large $\Lambda$ we are now actually free to allow $T_V$ to be as large as
particle physics suggests. Then, for such a large $T_V/T(t_0)$ we see that
the quantity $\bar{\Omega}_M(t_0)$ has to be completely negligible
today\footnote{$\bar{\Omega}_M(t_0)$ is suppressed by $G_{eff}$ being
small, and not by $\rho_{M}(t_0)$ itself being small, with $G_{eff}$ being
made smaller the larger rather than the smaller $S_0$ gets to be, to thus
enable the $c\Lambda/\rho_{M}(t_0)
=\bar{\Omega}_{\Lambda}(t_0)/\bar{\Omega}_M(t_0)$ ratio to be as large as
particle physics suggests while not leading to any 60 order of magnitude
conflict with observation.} so that $q_0$ must thus, without any
fine-tuning at all, necessarily lie between zero and minus one today
notwithstanding that $T_V$ is huge. The essence of the conformal gravity
approach then is not to change the matter and energy content of the
universe at all, but rather only to change their effect on cosmic
evolution, with $\Lambda$ itself no longer needing to be quenched.

In order to fit the Hubble plot data we need to determine the dependence of
$d_L$ on $z$ in the conformal theory, something we can readily do now that 
we have obtained the explicit form of the expansion factor $R(t)$. Thus, for
temperatures well below $T_{max}$ and for the naturally achievable
\cite{Mannheim2001a} $T_V \ll T_{max}$ case of most practical interest to
conformal gravity (viz. a case where $T_{max}^2T^2(t_0)/T_V^4$ can be of 
order one) we may set 
\begin{equation}
R(t)= (-k/\alpha)^{1/2} \sinh(\alpha^{1/2} ct)~~,
\label{9}
\end{equation}
so that 
\begin{equation}
-q_0=\tanh^2(\alpha^{1/2}ct_0)
=\alpha c^2/H^2_0~~,~~t_0= {\rm arctanh}
[(-q_0)^{1/2}]/(-q_0)^{1/2}H_0~~.
\label{10}
\end{equation}
For geodesics
$\int_{t_1}^{t_0}c dt/R(t)=\int_0^{r_1}dr /[1-kr^2]^{1/2}$ we thus obtain 
\begin{equation}
(-k)^{1/2}r_1
=\coth(\alpha^{1/2}ct_0)/\sinh(\alpha^{1/2}ct_1)-\coth(\alpha^{1/2}ct_1)/
\sinh(\alpha^{1/2}ct_0)~~.
\label{11} 
\end{equation}
Then, on noting that $\sinh(\alpha^{1/2}ct_1)=
(-q_0)^{1/2}/(1+q_0)^{1/2}(1+z)$ where $z=R(t_0)/R(t_1)-1$ and where $q_0$
is the current value of $q(t)$, we find that we can express the general
luminosity distance $d_L=r_1R(t_0)(1+z)$ entirely in terms of the current
era $H_0$ and $q_0$ according to the very compact relation
\cite{Mannheim2001b}
\begin{equation}
H_0 d_L/c=-(1+z)^2\left\{1-[1+q_0-q_0/(1+z)^2]^{1/2}\right\}/q_0~~.
\label{12}
\end{equation}
Conformal gravity fits to the luminosity distance can thus be parametrized 
via the one parameter $q_0$, a parameter which must lie somewhere between
zero and minus one, with $d_L$ thus having to lie somewhere between
$d_L(q_0=0)=cH_0^{-1}(z+z^2/2)$ and $d_L(q_0=-1)=cH_0^{-1}(z+z^2)$ at 
temperatures well below $T_{max}$. 

Having obtained Eq. (\ref{12}) we can now turn to a data analysis. On 
fitting the same 54 supernovae data points as previously, our best fit is
obtained for $q_0=-0.37$, $M=-19.37$ with $\chi^2=58.62$. We display this
fit as the upper curve in Fig. (1), and as we thus see, in the detected
region the best fits of the standard and conformal models are completely
indistinguishable, only in fact departing from each other at the very
highest available redshifts. For comparison purposes we find that for
$q_0=0$ a best fit value of  $\chi^2=61.49$ is obtained with
$M=-19.29$,\footnote{Such high quality fitting with $q_0=0$ has also been
noted by other authors \cite{Perlmutter1999,Dev2000} though not within the
context of conformal gravity.} with fits for other typical values of $q_0$
being reported in \cite{Mannheim2001b}.\footnote{If one also takes the
$\Delta z_i$ errors in the reported redshifts into consideration, the
conformal gravity chi squared values for the 54 points get reduced
\cite{Mannheim2001b} to $\chi^2(q_0=-0.33)=54.13$ and $\chi^2(q_0=0)=56.0$, 
while the standard model chi squared becomes
$\chi^2(\Omega_M(t_0)=0.3,\Omega_{\Lambda}(t_0)=0.7)=53.27$.}  Beyond the
purely phenomenological fact that the conformal gravity fits actually
provide a good accounting of the supernovae data at all, it is important
to stress that as such these fits are the first ones ever obtained in which
the cosmological constant is allowed to take a large unquenched particle
physics scale value, with the fits thus establishing the empirical fact
that it is in fact possible to fit the supernovae data without fine-tuning. 

With the standard cosmology requiring deceleration above $z=1$ and with the
conformal cosmology continuing to accelerate, extension of the Hubble plot
beyond $z=1$ will actually enable us to discriminate between the two 
cosmologies. We thus augment Fig. (2) by adding in the $z>1$ conformal
gravity predictions. The highest curve in Fig. (2) is the conformal gravity
prediction for $q_0=-0.37$, while the middle curve is that for $q_0=0$. As
we see, these two typical conformal gravity curves start to depart from the
standard model expectation fairly rapidly once $z>1$, with the three curves
in Fig. (2) respectively corresponding to apparent magnitudes $m=27.17$,
$m=27.04$ and $m=26.75$ at $z=2$, and to $m=30.40$, $m=30.25$ and $m=29.14$
at $z=5$. A quite modest extension of the Hubble plot will thus readily
enable us to discriminate between standard gravity and its conformal
alternative while potentially even being definitive for both.

\section {Outlook and challenges} 

While there has yet to be detailed exploration of the $z>1$ Hubble plot
using supernovae standard candles, it is of some interest to note that
recently a  first $z>1$ data point was actually established
\cite{Riess2001}, viz. the supernova SN 1997ff which was found to be at a
redshift $z=1.7^{+0.1}_{-0.15}$. To illustrate the data of \cite{Riess2001}
we have augmented Fig. (2) by adding in the 68\% and 95\% confidence region
values for the measured apparent magnitude $m$ at redshifts $z=1.65$,
$z=1.7$ and $z=1.75$. (In the figure the two inner horizontal bars on the
vertical data points represent the extent of the 68\% confidence region at
each of the chosen redshifts while the two outer bars represent the 95\%
confidence one.) While one should not read too much into a single data
point,\footnote{Indeed, a shortcoming of these particular data is that SN
1997ff just happens to be gravitationally lensed by two foreground galaxies
\cite{Riess2001}, with its "true" apparent magnitude thus likely to be
somewhat larger (viz. dimmer) than indicated in the figure.} it is
interesting to note that the data can accommodate both the standard and
conformal theories, with it being necessary to acquire a whole set of $z>1$
data points in order to identify any specific trend in the data that there
might be, with it as yet being too early to ascertain from available
supernovae data whether the $z>1$ universe is decelerating or accelerating.

Beyond the standard candle supernovae, it has also been noted by Daly
\cite{Daly2002} that the very powerful FRII bridge radio galaxies can serve 
as standard yardsticks, and can thus also be used to extract cosmological
parameters. As such, this technique serves to complement the supernovae
analyses, and even to potentially go beyond them since radio galaxies are
already being seen out to $z=2$ or so. Interestingly, the data presented in
\cite{Daly2002} are so far  found to be able to accommodate both standard
and conformal gravity, with further study of this issue thus having the
potential to be quite instructive. 

As we noted earlier, that apart from the $z\sim 1$ region Hubble plot,
cosmology  can also be tested at a variety of much larger redshifts, and it
is thus urgent to test conformal gravity at those redshifts as well. While
its predictions for the microwave background await the development of a
conformal cosmology galaxy fluctuation theory, its initial predictions for
nucleosynthesis have already been worked out \cite{Lohiya1999}. What was
found was that while the expanding and thus cooling conformal cosmology can
readily generate the requisite amounts of primordial helium and lithium,
because the cosmology expands altogether slower than the standard cosmology
its predictions for deuterium and for $^9Be$ are substantially different
from those of the standard model. Specifically, because of the slowness of
the expansion very little primordially generated deuterium manages to
survive, but because of this same slowness the cosmology is able to get
passed the $A=8$ nuclear fusion bottleneck (viz. the absence of any stable
nuclei with 8 nucleons) and thus primordially produce $^9Be$ and elements
heavier than it, in fact even producing $^9Be$ with its measured abundance,
an abundance 8 orders of magnitude greater than that generatable in the
standard theory. Now it was noted in \cite{Lohiya1999} that it would be
relatively easy to produce deuterium by spallation once inhomogeneities
begin to develop in the universe (i.e. post-primordial but pre-galactic).
With an explicit theory for such inhomogeneous deuterium production yet to
be developed, conformal gravity thus remains challenged by the deuterium
problem just as the standard theory remains challenged by the $^9Be$
problem. However, since conformal gravity so capably handles the most
vexing problem facing the standard theory, viz. the cosmological constant
problem, it would thus appear to merit further consideration. 

The author is indebted to Drs. R. A. Daly, G. V. Dunne, K. Horne, D. 
Lohiya, R. Plaga and B. E. Schaefer for helpful comments. This work has been
supported in  part by the Department of Energy under grant No.
DE-FG02-92ER40716.00.

\vfill\eject

\begin{figure} 
\epsfig{file=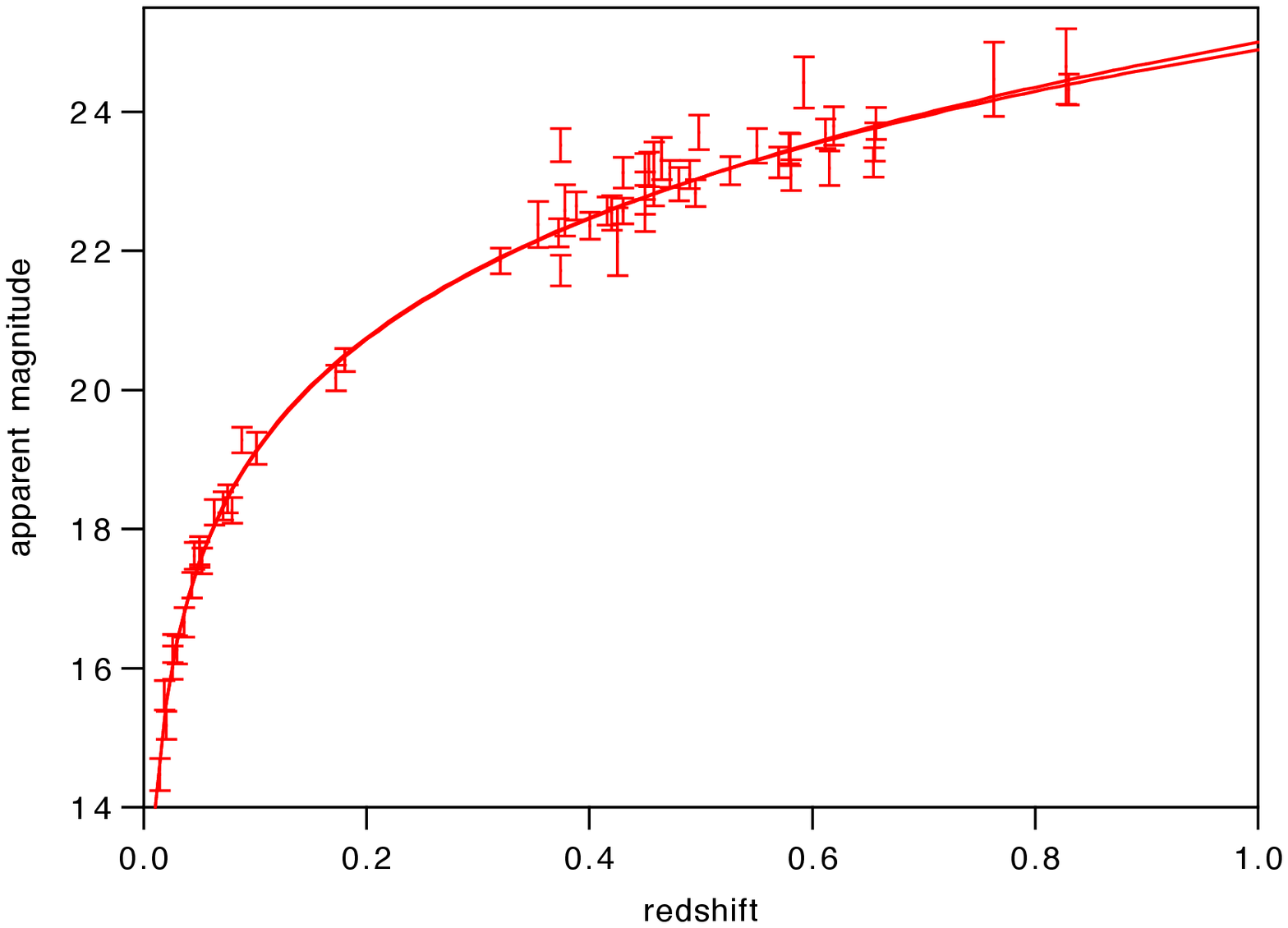,scale=0.75}
\caption{The $q_0=-0.37$ conformal gravity fit (upper curve) and the  
$\Omega_{M}(t_0)=0.3$, $\Omega_{\Lambda}(t_0)=0.7$ standard model fit (lower
curve) to the $z<1$ supernovae Hubble plot data.}
\label{Fig. (1)}
\end{figure}

\begin{figure}
\epsfig{file=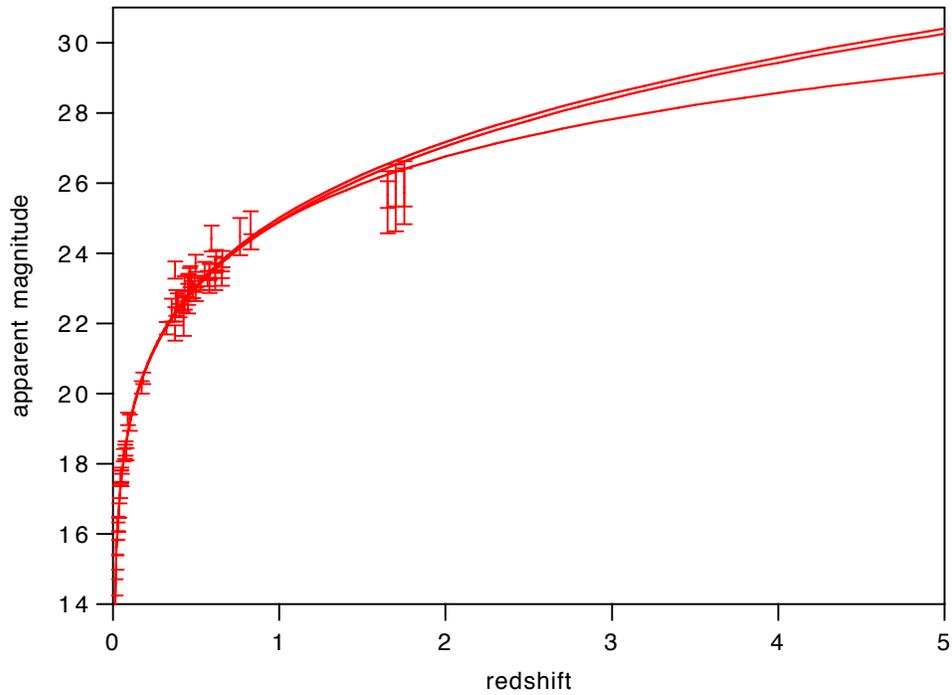,scale=0.75}
\caption{Hubble plot expectations for $q_0=-0.37$ (highest curve) and
$q_0=0$ (middle curve) conformal gravity and for
$\Omega_{M}(t_0)=0.3$,
$\Omega_{\Lambda}(t_0)=0.7$ standard gravity (lowest curve).}
\label{Fig. (2)}
\end{figure}

\end{document}